# CDDiff: Semantic Differencing for Class Diagrams

Shahar Maoz*, Jan Oliver Ringert**, and Bernhard Rumpe

Software Engineering
RWTH Aachen University, Germany
http://www.se-rwth.de/

**Abstract.** Class diagrams (CDs), which specify classes and the relationships between them, are widely used for modeling the structure of object-oriented systems. As models, programs, and systems evolve over time, during the development lifecycle and beyond it, effective change management is a major challenge in software development, which has attracted much research efforts in recent years.

In this paper we present *cddiff*, a semantic diff operator for CDs. Unlike most existing approaches to model comparison, which compare the concrete or the abstract syntax of two given diagrams and output a list of syntactical changes or edit operations, *cddiff* considers the *semantics* of the diagrams at hand and outputs a set of *diff witnesses*, each of which is an object model that is possible in the first CD and is not possible in the second. We motivate the use of *cddiff*, formally define it, and show how it is computed. The computation is based on a reduction to Alloy. The work is implemented in a prototype Eclipse plug-in. Examples show the unique contribution of our approach to the state-of-the-art in version comparison and evolution analysis.

## 1 Introduction

Class diagrams (CDs) are widely used for modeling the structure of object-oriented systems. The syntax of CDs includes classes and the various relationships between them (association, generalization, etc.). The semantics of CDs is given in terms of object models, consisting of sets of objects and the relationships between these objects. Specifically, we are interested in a variant of the standard UML2 CDs, which is rich and expressive, supporting generalizations (inheritance), interface implementation, abstract and singleton classes, class attributes, uni- and bi-directional associations with multiplicities, enumerations, aggregation, and composition.

As models, programs, and systems evolve over time, during the development lifecycle and beyond it, effective change management and controlled evolution are major challenges in software development, and thus have attracted much research

* S. Maoz acknowledges support from a postdoctoral Minerva Fellowship, funded by the German Federal Ministry for Education and Research.
** J.O. Ringert is supported by the DFG GK/1298 AlgoSyn.

efforts in recent years (see, e.g., [1, 5, 10, 14, 17, 22, 26, 34]). Fundamental building blocks for tracking the evolution of software artifacts are diff operators one can use to compare two versions of a program or a model. Most existing approaches to differencing concentrate on matching between model elements using different heuristics related to their names and structure and on finding and presenting differences at a concrete or abstract syntactic level. While showing some success, most of these approaches are also limited. Models that are syntactically very similar may induce very different semantics (in the sense of 'meaning' [12]), and vice versa, models that semantically describe the same system may have rather different syntactic representations. Thus, a list of syntactic differences, although accurate, correct, and complete, may not be able to reveal the real implications these differences may have on the correctness and potential use of the models involved. In other words, such a list, although easy to follow, understand, and manipulate (e.g., for merging), may not be able to expose and represent the semantic differences between two versions of a model, in terms of the bugs that were fixed or the features (and new bugs...) that were added.

In this paper we present *cddiff*, a semantic diff operator for CDs. Unlike existing differencing approaches, *cddiff* is a *semantic diff operator*. Rather than comparing the concrete or the abstract syntax of two given diagrams, and outputting a list of syntactical changes or edit operations, *cddiff* considers the semantics of the diagrams at hand and outputs a set of *diff witnesses*, each of which is an object model that is possible in the first CD and is not possible in the second. These object models provide concrete proofs for the meaning of the change that has been done between the two compared versions and for its effect on the use of the models at hand.

We specify CDs using the class diagrams of UML/P [29], a conceptually refined and simplified variant of UML designed for low-level design and implementation. Our semantics of CDs is based on [11] and is given in terms of sets of objects and relationships between these objects. An overview of the formal definition of the syntax and semantics of our CDs is given in Sect. 3.

Given two CDs, $cd_1$ and $cd_2$, $cddiff(cd_1, cd_2)$ is roughly defined as the set of object models possible in the first CD and not possible in the second. As this set may be infinite, we are specifically interested in its bounded version, $cddiff_k(cd_1, cd_2)$, which only includes object models where the number of object instances is not greater than $k$. The formal definition of *cddiff* is given in Sect. 4.

To compute *cddiff* we use Alloy [13]. Alloy is a textual modeling language based on relational first-order logic. A short overview of Alloy is given in Sect. 3.2. To employ Alloy for our needs, we have defined a transformation that takes two CDs and generates a single Alloy module. The module includes predicates specifying each of the CDs, `cd1` and `cd2`, and a diff predicate reading `Cd1NotCd2`, specifying the existence of a satisfying assignment for the predicate `cd1` (for us, representing an instance of $cd_1$), which is not a satisfying assignment for the predicate `cd2` (representing an instance of $cd_2$). Analyzing this predicate with a user-specified scope $k$ produces elements of $cddiff_k(cd_1, cd_2)$, if any, as required.

Our transformation is very different from ones suggested in other works that use Alloy to analyze CDs (see, e.g., [4, 21]). First, we take two CDs as input, and output one Alloy module. Second, to support a comparison in the presence of generalizations and associations, we must use a non-shallow embedding, that is we have to encode all the relationships between the signatures in generated predicates ourselves. In particular, we cannot use Alloy's `extends` keyword to model inheritance and cannot use Alloy's fields to model class attributes, because for the shared signatures, a class's inheritance relation and set of attributes may be different between the two CDs. Thus, these need to be modeled as predicates, different ones for each CD, outside the signatures themselves. The transformation is described in Sect. 4.2.

In addition to finding concrete diff witnesses (if any exist), which demonstrate the meaning of the changes that were made between one version and another, *cddiff* can be used to compare two CDs and decide whether one CD semantics includes the other CD semantics (the latter is a refinement of the former), are they semantically equivalent, or are they semantically incomparable (each allows instantiations that the other does not allow). When applied to the version history of a certain CD, which can be retrieved from a version repository, such an analysis provides a semantic insight into the evolution of this CD, which is not available in existing syntactic approaches.

We have implemented *cddiff* and integrated it into a prototype Eclipse plug-in. The plug-in allows the engineer to compare two selected CDs and to browse the diff witnesses found, if any. Indeed, all examples shown in this paper have been computed by our plug-in. We describe the plug-in's implementation, main features, and performance results in Sect. 5.

Following the evaluation in Sect. 5, we define and implement two important extensions of the basic *cddiff* technique. The first extension deals with filtering the diff witnesses found, so that 'uninteresting witnesses' are filtered out, and a more succinct yet informative set of witnesses is provided to the engineer. The second extension deals with the use of abstraction in the comparison. The extensions are described in Sect. 6.

Model and program differencing, in the context of software evolution, has attracted much research efforts in recent years (see [1, 5, 10, 14, 17, 22, 26, 34]). In contrast to our work, however, most studies in this area present syntactic differencing, at either the concrete or the abstract syntax level. We discuss related work in Sect. 8.

It is important not to confuse differencing with merging. Merging is a very important problem, dealing with reconciling the differences between two models that have evolved independently from a single source model, by different developers, and now need to be merged back into a single model (see, e.g., [3, 10, 17, 23, 25]). Differencing, however, is the problem of identifying the differences between two versions, for example, an old version and a new one, so as to better understand the course of a model evolution during some step of its development. Thus, diff witnesses are not conflicts that need to be reconciled. Rather, they are proofs of features that were added or bugs that have been fixed from one version to another along the history of the design and development process.

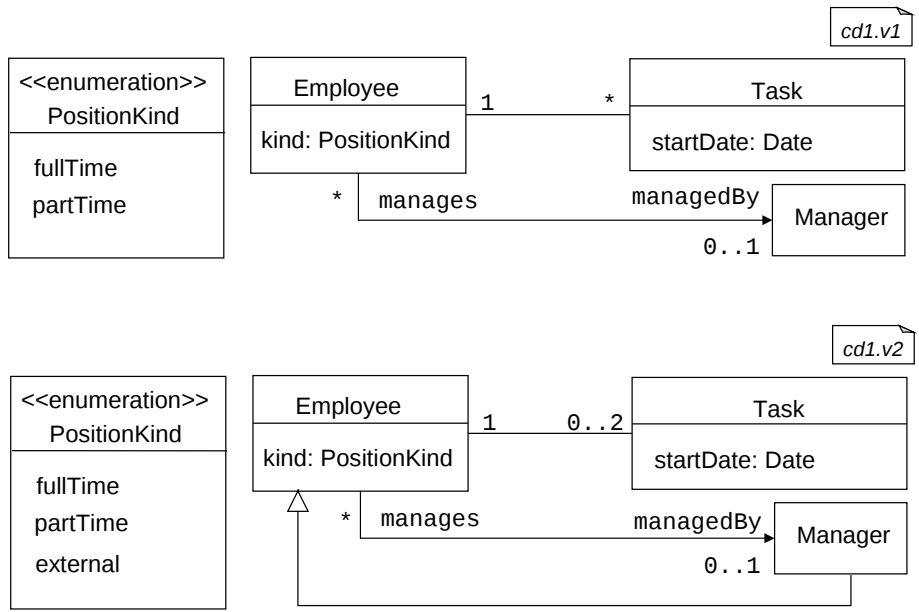

**Fig. 1.** $cd1.v1$ and its revised version $cd1.v2$

The next section presents motivating examples demonstrating the unique features of our work. Sect. 3 provides preliminary definitions of the CD language syntax and semantics as used in our work. Sect. 4 introduces *cddiff* and the technique to compute it. Sect. 5 presents the prototype implementation and related applications. Sect. 6 describes the filtering and abstraction extensions. Sect. 7 presents a discussion of advanced topics and future work directions, Sect. 8 considers related work, and Sect. 9 concludes.

## 2 Examples

We start off with motivating examples for semantic differencing of CDs. The examples are presented semi-formally. Formal definitions appear in Sect. 4.

### 2.1 Example I

Consider $cd1.v1$ of Fig. 1, describing a first version of a model for (part of) a company structure with employees, managers, and tasks. A design review with a domain expert has revealed three bugs in this model: first, employees should not be assigned more than two tasks; second, managers are also employees, and they can handle tasks too; and third, there is another kind of position, namely an external position.

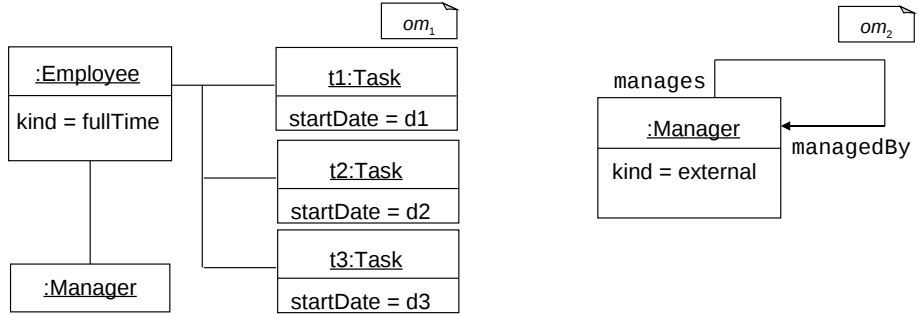

**Fig. 2.** Example object models representing semantic differences between the old class diagram $cd1.v1$ and its revised version $cd1.v2$.

Following this design review, the engineers created a new version $cd1.v2$, shown in the same figure. The two versions share the same set of named elements but they are not identical. Syntactically, the engineers added an inheritance relation between Manager and Employee, set the multiplicity on the association between Employee and Task to 0..2, and added the external position kind. What are the semantic consequences of these differences?

Using the operator *cddiff* we can answer this question. $cddiff(cd1.v1, cd1.v2)$ outputs $om_1$, shown in Fig. 2, as a diff witness that is in the semantics of $cd1.v1$ and not in the semantics of $cd1.v2$; thus, it demonstrates (though does not prove) that the bug of having more than two tasks per employee was fixed. In addition, $cddiff(cd1.v2, cd1.v1)$ outputs $om_2$, shown in Fig. 2 too. $om_2$ is a diff witness that is in the semantics of the new version $cd1.v2$ and not in the semantics of the old version $cd1.v1$. Thus, the engineers should perhaps check with the domain expert whether the model should indeed allow managers to manage themselves and hold an external kind of position.

### 2.2 Example II

The two class diagrams $cd3.v1$ and $cd3.v2$, shown in Fig. 3, provide alternative descriptions for the relation between Department and Employee in the company. Again the two diagrams share the same set of named elements but the diagrams are not identical. First, Department is a singleton only in $cd3.v1$. Second, only in $cd3.v1$ the relation between Department and Employee is a Whole/Part composition relation. What are the semantic consequences of the differences between the two versions of $cd3$?

Fig. 3 includes two objects models. In $om_3$ there are two departments with no employees. In $om_4$ there is a single employee and no departments. It is easy to see that both object models are in the semantics of $cd3.v2$ but not in the semantics of $cd3.v1$. We formally write it as $\{om_3, om_4\} \subseteq cddiff(cd3.v2, cd3.v1)$. In addition, we can see that $cd3.v2$ is a refinement of $cd3.v1$, since all object models in the semantics of $cd3.v1$ are also in the semantics of $cd3.v2$ (that is, $cddiff(cd3.v1, cd3.c2) = \emptyset$). Again, the two diff witnesses (in one direction) can be computed and the refinement relation (in the other direction) can be proved (in a bounded scope) by our operator.

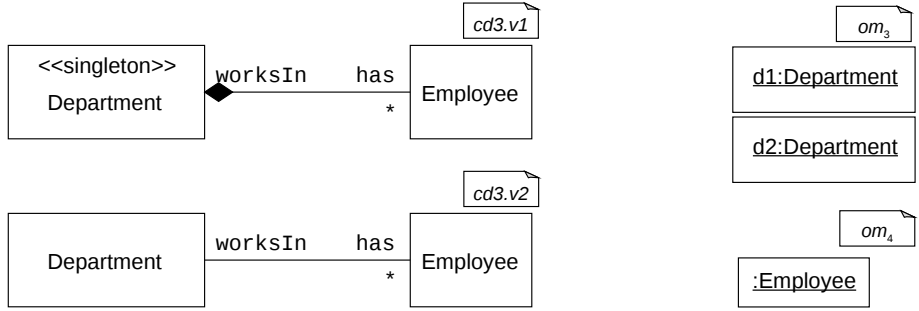

**Fig. 3.** $cd3.v1$ and its revised version $cd3.v2$, with example object models representing the semantic differences between them. Both object models are in the semantics of $cd3.v2$ and not in the semantics of $cd3.v1$.

### 2.3 Example III

Finally, $cd5.v1$ of Fig. 4 is another class diagram from this model of company structure. In the process of model quality improvement, an engineer has suggested to refactor it by introducing an abstract class `Person`, replacing the association between `Employee` and `Address` by an association between `Person` and `Address`, and redefining `Employee` to be a subclass of `Person`. The resulting suggested CD is $cd5.v2$.

Using *cddiff* we are able to prove (in a bounded scope) that despite the syntactic differences, the semantics of the new version is equivalent to the semantics of the old one, formally written $cddiff(cd5.v1, cd5.v2) = cddiff(cd5.v2, cd5.v1) = \emptyset$. The refactoring is correct and the new suggested version can be committed.

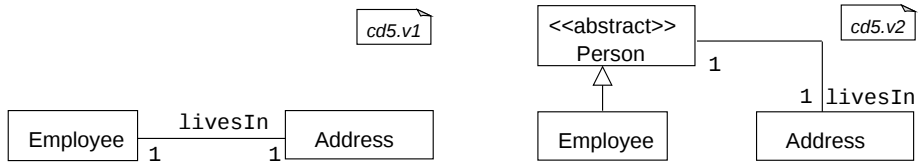

**Fig. 4.** $cd5.v1$ and its revised version $cd5.v2$. The two versions have equal semantics.

## 3 Preliminaries

We give a short overview of the CD language used in our work and of Alloy, the tool we use for the computation of *cddiff*.

### 3.1 Class diagrams language

As a concrete CD language we use the class diagrams of UML/P [29], a conceptually refined and simplified variant of UML designed for low-level design and implementation. Our semantics of CDs is based on [11] and is given in terms of sets of objects and relationships between these objects. More formally, the semantics is defined using three parts: a precise definition of the syntactic domain, i.e., the syntax of the modeling language CD and its context conditions (we use MontiCore [16, 24] for this); a semantic domain - for us, a subset of the System Model (see [7, 8]) OM, consisting of all finite object models; and a mapping $sem : CD \rightarrow \mathcal{P}(OM)$, which relates each syntactically well-formed CD to a set of constructs in the semantic domain OM. For a thorough and formal account of the semantics see [8].

Note that we use a *complete* interpretation for CDs (see [29] ch. 3.4). This roughly means that 'whatever is not in the CD, should indeed not be present in the object model'. In particular, we assume that the list of attributes of each class is complete, e.g., an `employee` object with an `id` and a `salary` is not considered as part of the semantics of an `Employee` class with an `id` only.

The CD language constructs we support include generalization (inheritance), interface implementation, abstract and singleton classes, class attributes, uni- and bi-directional associations with multiplicities, enumerations, aggregation, and composition.

### 3.2 A brief overview of Alloy

Alloy [2, 13] is a textual modeling language based on relational first-order logic. An Alloy module consists of a number of signature declarations, fields, facts and predicates. The basic entities in Alloy are atoms. Each signature denotes a set of atoms. Each field belongs to a signature and represents a relation between two or more signatures. Such relations are interpreted as sets of tuples of atoms. Facts are statements that define constraints on the elements of the module. Predicates are parametrized constraints, which can be included in other predicates or facts.

Alloy Analyzer is a fully automated constraint solver for Alloy modules. The analysis is achieved by an automated translation of the module into a Boolean expression, which is analyzed by SAT solvers embedded within the Analyzer. The analysis is based on an exhaustive search for instances of the module. The search space is bounded by a user-specified scope, a positive integer that limits the number of atoms for each signature in an instance of the system that the solver analyzes.

The Analyzer can check for the validity of user-specified assertions. If an instance that violates the assertion is found within the scope, the assertion is not valid. If no instance is found, the assertion might be invalid in a larger scope. Used in the opposite way, one can look for instances of user-specified predicates. If the predicate is satisfiable within the given scope, the Analyzer will find an instance that proves it. If not, the predicate may be satisfiable in a larger scope. We discuss the advantages and limitations of using Alloy for our problem in Sect. 7. A thorough account of Alloy can be found in [13].

## 4 CDDiff

### 4.1 Definitions

We define a diff operator $cddiff : CD \times CD \to \mathcal{P}(OM)$, which maps two CDs, $cd_1$ and $cd_2$, to the (possibly infinite) set of all object models that are in the semantics of $cd_1$ and are not in the semantics of $cd_2$. Formally:

**Definition 1.** $cddiff(cd_1, cd_2) = \{om \in OM \mid om \in sem(cd_1) \land om \notin sem(cd_2)\}$.

Note that $cddiff$ is not symmetric. In addition, by definition, $\forall cd_1, cd_2 \in CD$, $cddiff(cd_1, cd_1) = \emptyset$ (the empty set, not the empty object model) and $cddiff(cd_1, cd_2) \cap cddiff(cd_2, cd_1) = \emptyset$, as expected. The members of the set $cddiff$ are called *diff witnesses*.

The set-theoretic definition of *cddiff*, as given above, is however not constructive, and may yield an infinite set. As a pragmatic solution, we approximate it by defining (a family of) bounded diff operators that we are able to compute. Thus, we use a bound $k$, which limits the total number of objects in the diff witnesses we are looking for. Formally:

**Definition 2.** $\forall k \geq 0, cddiff_k(cd_1, cd_2) = \{om \mid om \in cddiff(cd_1, cd_2) \land |om| \leq k\}$, where $|om|$ is the total number of objects in $om$.

### 4.2 Computing $cddiff_k$: overview

To compute $cddiff_k$ we use Alloy. To employ Alloy for our needs, we have defined a transformation that takes two CDs and generates a single Alloy module. The module includes predicates specifying each of the CDs, `cd1` and `cd2`, and a diff predicate reading `Cd1NotCd2`, specifying the existence of a satisfying assignment for the predicate `cd1` (for us, representing an instance of $cd_1$), which is not a satisfying assignment for the predicate `cd2` (representing an instance of $cd_2$). Analyzing this predicate with a user-specified scope $k$ produces elements of $cddiff_k(cd_1, cd_2)$, if any, as required.

Our transformation is very different from ones suggested in other works that use Alloy to analyze CDs (see, e.g., [4, 21]). First, we take two CDs as input, and output a single Alloy module. Second, to support a comparison in the presence of generalizations and associations, we must use a non-shallow embedding, that is we have to encode all the relationships between the signatures in generated predicates ourselves. In particular, we cannot use Alloy's `extends` keyword to model inheritance and cannot use Alloy's fields to model class attributes, because for the shared signatures, a class's inheritance relation and set of attributes may be different between the two CDs. Thus, these need to be modeled as predicates, different ones for each CD, outside the signatures themselves.

It is important to note that a naive approach that would transform each of the two CDs separately into a corresponding Alloy module and then compare the instances found by the analyzer for each CD, would have been incomplete and hopelessly inefficient. Such an approach requires the complete computation of the two sets of instances before the comparison could be done. As Alloy generates instances one-by-one, with no guarantee about their order, this could not work in practice. Thus, our approach, of taking the two input CDs and constructing a single Alloy module whose all instances, if any, are diff witnesses, is indeed required. In other words, instead of computing the differences, if any, ourselves, we create an Alloy module whose instances are the differences we are looking for, and let the SAT solver do the hard work for us.

Below we show only selected excerpts from the generated Alloy module corresponding to the two CDs from the example in Fig. 1 (a complete definition of the translation, which shows how each CD construct is handled, is given in supporting materials available from [31]).

### 4.3 Computing $cddiff_k$: the generated Alloy module

We start off with a generic part, which is common to all our generated modules.

```
// Names of fields/associations in classes of the model
abstract sig FName {}

// Parent of all classes relating fields and values
abstract sig Obj { get: FName -> {Obj + Val + EnumVal}}

// Values of fields
abstract sig Val {}

fact values {
  // No values can exist on their own
  all v: Val | some f: FName | v in Obj.get[f] }

//Names of enum values in enums of the model
abstract sig EnumVal {}

fact enums {
  //no enum values can exist on their own
  all v: EnumVal | some f: FName | v in Obj.get[f] }
```

**Listing 1.1.** `FName`, `Obj`, `Val`, and `EnumVal` signatures

List. 1.1 shows the abstract signature `FName` used to represent association role names and attribute names for all classes in the module. The abstract `Obj` signature is the parent of all classes in the module, and its `get` Alloy field relates it and an `FName` to instances of `Obj`, `Val`, and `EnumVal`. List. 1.1 also shows the abstract signature `Val`, which we use to represent all predefined types (i.e., primitive types and other types that are not defined as classes in the CDs). Values of enumeration types are represented using signature `EnumVal`. Enumeration values as well as primitive values should only appear in an instance if referenced by any object (see predicates in lines 10-12 and lines 17-19).

```
1  pred ObjAttrib[objs:set Obj,
2          fName:one FName, fType:set {Obj + Val + EnumVal}] {
3    objs.get[fName] in fType
4    all o: objs| one o.get[fName] }
5
6  pred ObjNoFName[objs:set Obj, fName:set FName] {
7    no objs.get[fName] }
```

**Listing 1.2.** Predicates for objects and their fields

List. 1.2 shows some of the generated predicates responsible for specifying the relation between objects and fields: `ObjAttrib` limits `objs.get[fName]` to the correct field's type and ensures that there is exactly one atom related to the field name (by the `get` relation); `ObjNoFName` is used to ensure classes do not have field names other than the ones stated in the CD.

```
1  pred ObjUAttrib[objs:set Obj,
2       fName:one FName, fType:set Obj, up: Int] {
3    objs.get[fName] in fType
4    all o: objs| (#o.get[fName] =< up) }
5
6  pred Composition[left:set Obj,
7          lFName:one FName, right:set Obj] {
8    all l1, l2: left |
9      (# {l1.get[lFName] & l2.get[lFName]} > 0) => l1=l2
10   all r: right | # {l: left | r in l.get[lFName]} = 1 }
```

**Listing 1.3.** Predicates for multiplicities and Whole/Part compositions

List. 1.3 shows some of the generated predicates responsible to specify multiplicities and Whole/Part compositions. The first predicate provides an upper bound for the number of objects in the set represented by the `get` relation for a specified role name. The second predicate is used to constrain a composition relation between classes. Its first statement (lines 8-9) ensures that no two wholes (on the 'left') own the same part (on the 'right'). The second statement (line 10) ensures that a part (on the 'right') belongs to exactly one whole (on the 'left').

```
1  // Predicate for diff
2  pred Cd1NotCd2 { cd1 not cd2}
3
4  // Command for diff
5  run Cd1NotCd2 for 5
```

**Listing 1.4.** The diff predicate and the related run command

List. 1.4 shows the simple predicate `Cd1NotCd2` representing the diff. An Alloy instance that satisfies the generated predicate `cd1` and does not satisfy the generated predicate `cd2` is in the set $cddiff_k(cd_1, cd_2)$. The value for the scope $k$ of the `run` command (line 5) is part of the input of our transformation.

All the above are generic, that is, they are common to all generated modules, independent of the input CDs at hand. We now move to the parts that are specific to the two input CDs.

All class names and field names from the two CDs are shown in List. 1.5 as Alloy signatures and are stripped from their inheritance relations, attributes, associations etc. Note the `type_Date` signature in line 6, which extends `Val` (see List. 1.1). Concrete enumeration values from both class diagrams are declared in lines 9-10.

```
// Concrete names of fields in cd1 and cd2
one sig startDate, mngBy, worksOn, mng,
        doneBy, kind extends FName {}

// Concrete value types in model cd1 and cd2
lone sig type_Date extends Val {}

// Concrete enum values
lone sig enum_PosKnd_external, enum_PosKnd_fullTime,
         enum_PosKnd_partTime extends EnumVal {}

// Actual classes in the model
sig Tsk, Emp, Mgr extends Obj {}
```

**Listing 1.5.** The common signatures

Next, we define a set of functions and a predicate for each CD individually. We show here only the ones for $cd1.v2$, a CD that we presented in Fig. 1 (in the generated Alloy code that we show below, this CD appears as `cd2`).

First, subtype functions, shown in List. 1.6 (top), which specify subtype relations between the relevant signatures specific for this CD. Note how function `EmpSubsCD2` denotes that in $cd1.v2$ employees are either of type `Emp` or their subtype `Mgr`. The possible values of enumeration `PosKnd` in $cd1.v2$ are defined by function `PosKndEnumCD2`.

Second and finally, the predicate `cd2`, specifying the properties of the CD $cd1.v2$, is shown in List. 1.6. Note the use of the generic predicates defined earlier, in particular, the use of the parametrized predicate `ObjNoFName` (defined in List. 1.2); e.g., line 15 specifies that a `Tsk` has no other field names but `doneBy` and `startDate`. Also, note the use of the parametrized predicate `ObjLUAttrib` (defined using the predicate shown in List. 1.3); e.g., line 27 specifies that all instances of `Emp` (including subtypes, see the function `EmpSubsCD2` defined in List.1.6), work on at most 2 tasks.

```
1  // Types wrapping subtypes
2  fun MgrSubsCD2: set Obj { Mgr}
3  fun TskSubsCD2: set Obj { Tsk}
4  fun EmpSubsCD2: set Obj { Mgr + Emp}
5
6  // Enums
7  fun PosKndEnumCD2: set EnumVal { enum_PosKnd_external +
8    enum_PosKnd_fullTime + enum_PosKnd_partTime }
9
10 // Values and relations in cd2
11 pred cd2 {
12
13   // Definition of class Tsk
14   ObjAttrib[Tsk, startDate, type_Date]
15   ObjNoFName[Tsk, FName - doneBy - startDate]
16
17   // Definition of class Emp
18   ObjAttrib[Emp, kind, PosKndEnumCD2]
19   ObjNoFName[Emp, FName - kind - mngBy - worksOn]
20
21   // Definition of class Mgr
22   ObjAttrib[Mgr, kind, PosKndEnumCD2]
23   ObjNoFName[Mgr, FName - kind - mngBy - worksOn]
24
25   // Associations
26   BidiAssoc[EmpSubsCD2, worksOn, TskSubsCD2, doneBy]
27   ObjLUAttrib[EmpSubsCD2, worksOn, TskSubsCD2, 0, 2]
28   ObjLUAttrib[TskSubsCD2, doneBy, EmpSubsCD2, 1, 1]
29
30   ObjLUAttrib[EmpSubsCD2, mngBy, MgrSubsCD2, 0, 1]
31   ObjL[MgrSubsCD2, mngBy, EmpSubsCD2, 0] }
```

**Listing 1.6.** Subtyping functions and the predicate for $cd1.v2$

As an optional optimization, the transformation identifies and ignores syntactically equal attributes of same-name classes and common enumeration values between the two CDs. By definition, such attributes and enumerations will not be a necessary part of any diff witness and thus they can be ignored. Note that this is done on the flattened model, that is, while considering also inherited attributes. In addition to faster performance, this has the very important effect of reducing the size of the problem for Alloy, and hence, let us increase the maximum number of instances – Alloy's scope – in finding a witness, while keeping the size of the SAT problem small, and thus better cope with the bounded analysis limitation. In particular, in the presence of large CDs, it allows us to find differences that we were unable to find otherwise.

## 5 Implementation and Evaluation

We have implemented *cddiff* and integrated it into a prototype Eclipse plug-in. The input for the implementation are UML/P CDs, textually specified using MontiCore grammar and generated Eclipse editor [16, 24]. The plug-in transforms the input CDs into an Alloy module and uses Alloy's APIs to analyze it and to produce diff witnesses. Witnesses are presented to the engineer using MontiCore object diagrams. The complete analysis cycle, from parsing the two selected CDs, to building the input for Alloy, to running Alloy, and to translating the Alloy instances that were found, if any, back to MontiCore object diagrams, is fully automated.

### 5.1 Browsing diff witnesses

The plug-in allows the engineer to compare two selected CDs, and to browse the diff witnesses found, if any. Fig. 5 shows an example screen capture, where the engineer has selected to compare $cd1.v1$ (left) and $cd1.v2$ (right), which we presented in Sect. 2, and is currently browsing one of the two diff witnesses that were found. This witness is an object diagram with a full-time employee handling three tasks.

Clicking `Compute` computes the diff witnesses and shows a message telling the engineer if any were found. The diff witness is textually displayed as an object diagram in the central lower pane. The `Next` and `Previous` buttons browse for the next and previous diff witnesses. The `Switch Left/Right` button switches the order of comparison. The `Settings` button opens a dialog that allows the engineer to set values for several parameters, such as the scope that Alloy should use in the computation and the activation of various filters and abstractions (see Sect. 6).

### 5.2 High-level evolution analysis

Another application enabled by the plug-in is high-level evolution analysis. The plug-in supports a `compare` command: given two CDs, $cd_1$ and $cd_2$, and a scope $k$, the command checks whether one CD is a refinement of the other, are the two CDs semantically equivalent, or are they semantically incomparable (each allows object models the other does not allow). Formally, $compare(cd_1, cd_2, k)$ returns one of four answers:

$$
\begin{array}{ll}
<_k & \text{if } cddiff_k(cd_1, cd_2) = \emptyset \text{ and } cddiff_k(cd_2, cd_1) \neq \emptyset \\
>_k & \text{if } cddiff_k(cd_1, cd_2) \neq \emptyset \text{ and } cddiff_k(cd_2, cd_1) = \emptyset \\
\equiv_k & \text{if } cddiff_k(cd_1, cd_2) = \emptyset \text{ and } cddiff_k(cd_2, cd_1) = \emptyset \\
<>_k & \text{if } cddiff_k(cd_1, cd_2) \neq \emptyset \text{ and } cddiff_k(cd_2, cd_1) \neq \emptyset
\end{array}
$$

The subscript $k$ denotes the scope used in the computation.

Given a reference to a series of historical versions of a CD, as can be retrieved from the CD's entry in a revision repository (such as SVN, CVS etc.), the plug-in can use the `compare` command to compute a high-level analysis of the evolution of the CD: which new versions have introduced new possible implementations relative to their predecessors, which new versions have eliminated possible implementations relative to their predecessors, and which new versions included only syntactical changes that have not changed the semantics of the CD.

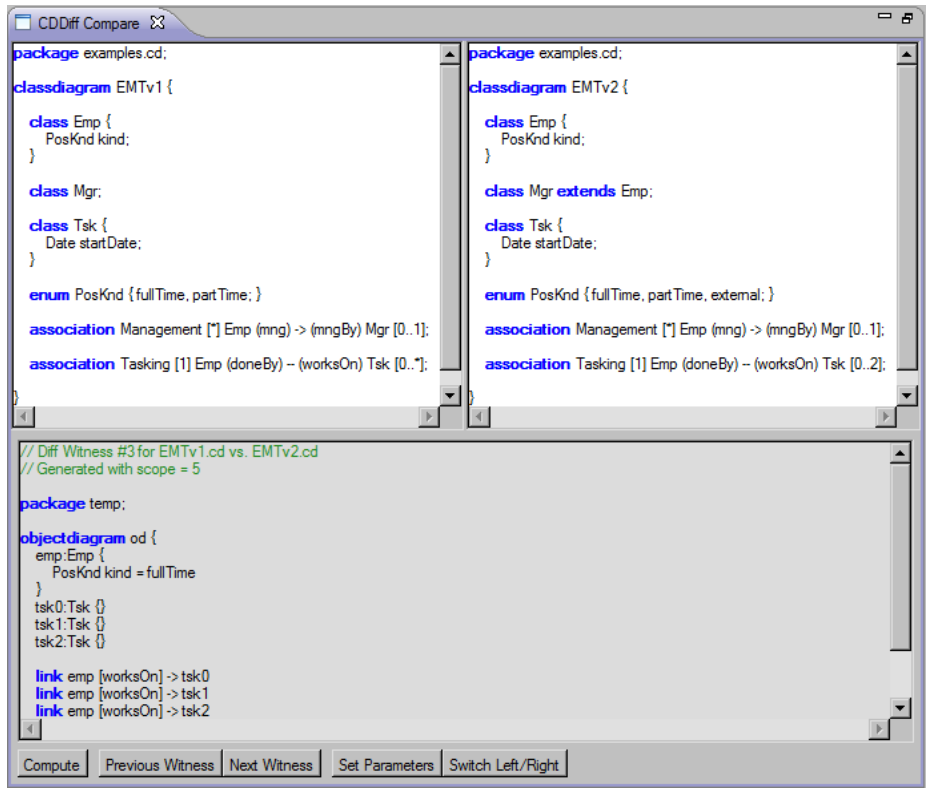

**Fig. 5.** A screen capture from Eclipse, showing a view from the prototype plug-in for *cddiff*. Two class diagrams that were selected by the user, corresponding to $cd1.v1$ and $cd1.v2$ of Fig. 1, are shown at the upper part of the screen. A generated diff witness, consisting of an object model that includes a full-time employee with three tasks, is displayed at the lower part of the screen.

For example, applying this evolution analysis to the examples presented in Sect. 2 with, e.g., a scope of 5, reveals: $compare(cd1.v1, cd1.v2, 5) = <>_5$, $compare(cd3.v1, cd3.v2, 5) = <_5$, and $compare(cd5.v1, cd5.v2, 5) = \equiv_5$. Thus, it shows (within the selected scope), that $cd1.v1$ and $cd1.v2$ are incomparable (each allows object models that are not allowed by the other), that $cd3.v1$ is a refinement of $cd3.v2$ (the latter allows all the object models that are allowed by the former, and some more), and that $cd5.v1$ and $cd5.v2$ have equivalent semantics (one is a correct refactoring of the other).

### 5.3 Performance

We report the performance of the plug-in in generating diff witnesses. Experiments were done using Alloy version 4.1.10 with SAT4J [30], on a laptop computer, Intel Dual Core CPU, 2.8 GHz, with 4 GB RAM, running Windows Vista.

Table 1 shows results from computing diff witnesses for the three examples presented in Sect. 2 using different scopes. Each example is reported twice, computing the differences in both directions. The column titled `Vars / primary vars / clauses` reports on the SAT formula created by Alloy. The column titled `Alloy time` reports the time it took Alloy to find the first diff witness (building the formula + finding the instance). The column titled `# Witnesses` reports on the total number of witnesses found by the plug-in (we compute only the first 20 witnesses). The rightmost column reports the total time it took for the plug-in to compute all (up to 20) witnesses. All timing data is reported in milliseconds.

Table 2 shows the results from computing diff witnesses for several versions of CDs from a library example (The CDs of the library example are available for download as supporting materials in [31]). The CDs in this example include 11 classes, 4 enumerations (with average of 4 values each), 7 associations (with most multiplicities ∗ or 1..∗), an average of 4 attributes per class (some classes have 6 attributes), and an inheritance hierarchy of depth 3. The columns in Table 2 are the same as the ones in Table 1.

On the one hand, the performance results show that for relatively small models, computing diff witnesses using our approach runs very fast or at least in reasonable times. On the other hand, the results show that for large models, or ones that require a high scope, performance may not scale well, as doubling the scope typically causes a performance slowdown of a factor of 4 or more. Given these results, in the future, we plan to develop heuristics to improve the scalability of *cddiff*, using, e.g., abstraction / refinement techniques, decomposition for early detection of independent sub models, etc. See the short discussions in the next section.

Finally, as can be seen from the table, in all cases where witnesses exist, the plug-in has found 20 witnesses (and could have perhaps found more if we would have continued to look for more witnesses). This points to a limitation in *cddiff*, where despite the symmetry breaking heuristics employed by Alloy, many of the witnesses found are rather similar and thus not interesting. To address this limitation, we have defined and implemented a filtering mechanism. We discuss this in Sect. 6.1.

## 6 Extensions: Filtering and Abstraction

### 6.1 Filtering diff witnesses

One limitation of *cddiff* and its computation through Alloy as presented in previous sections, is related to the usefulness of the set of witnesses that we find. In some cases, the automatically generated set contains many very similar and thus possibly uninteresting witnesses. This is true despite the symmetry reduction heuristics employed by Alloy. For example, assuming a difference in multiplicities of $*$ and $0..m$ between employees and tasks, all object models with one or more employees, where at least one employee has more than $m$ tasks are diff witnesses. Indeed, all such witnesses (up to the specified scope) may be returned by our computation. Thus, we look for ways to improve the usefulness of the computation by filtering out 'uninteresting witnesses' and keeping a more succinct yet informative set of witnesses.

| Name | Scope | Vars/p. vars/clauses | Alloy time (ms) | # Wit. | Plug-in time (ms) |
|---|---|---|---|---|---|
| Ex. 1 | 5 | 4079 / 234 / 9106 | 54 + 11 | 20 | 281 |
| Ex. 1 rev. | 5 | 4079 / 234 / 9092 | 44 + 7 | 20 | 212 |
| Ex. 1 | 10 | 13664 / 704 / 33386 | 265 + 29 | 20 | 634 |
| Ex. 1 rev. | 10 | 13664 / 704 / 33357 | 253 + 20 | 20 | 603 |
| Ex. 1 | 20 | 49834 / 2394 / 126446 | 1740 + 156 | 20 | 3472 |
| Ex. 1 rev. | 20 | 49834 / 2394 / 126387 | 1786 + 112 | 20 | 2970 |
| Ex. 2 | 5 | 883 / 72 / 2014 | 8 + 2 | 0 | 11 |
| Ex. 2 rev. | 5 | 883 / 72 / 2014 | 8 + 1 | 20 | 76 |
| Ex. 2 | 10 | 3613 / 242 / 9244 | 40 + 2 | 0 | 44 |
| Ex. 2 rev. | 10 | 3613 / 242 / 9244 | 40 + 5 | 20 | 164 |
| Ex. 2 | 20 | 14713 / 882 / 39484 | 337 + 10 | 0 | 348 |
| Ex. 2 rev. | 20 | 14713 / 882 / 39484 | 347 + 18 | 20 | 814 |
| Ex. 3 | 5 | 1165 / 77 / 2665 | 10 + 3 | 0 | 14 |
| Ex. 3 rev. | 5 | 1165 / 77 / 2665 | 10 + 3 | 0 | 14 |
| Ex. 3 | 10 | 4455 / 252 / 11020 | 56 + 28 | 0 | 84 |
| Ex. 3 rev. | 10 | 4455 / 252 / 11020 | 49 + 20 | 0 | 70 |
| Ex. 3 | 20 | 17575 / 902 / 45010 | 390 + 388 | 0 | 780 |
| Ex. 3 rev. | 20 | 17575 / 902 / 45010 | 397 + 404 | 0 | 802 |

**Table 1.** Results from computing diff witnesses for the three examples presented in Sect. 2, using different scopes. Each example is reported twice, computing the differences in both directions. The column titled `Vars / p. vars / clauses` reports on the SAT formula created by Alloy. The column titled `Alloy time` reports the time it took Alloy to find the first diff witness (building the formula + finding the instance). The column titled `# Wit.` reports on the total number of witnesses found by the plug-in (we compute up to 20 witnesses). The rightmost column reports the total time it took for the plug-in to compute all (up to 20) witnesses. All timing data is reported in milliseconds.

To address this problem, we have defined and implemented a filtering mechanism. At every stage of the computation, given the set of witnesses that was already found, the mechanism supports the filtering of witnesses that (1) only include objects of classes instantiated in previously found witnesses (`NNC`), (2) only include types of associations appearing in previously found witnesses (`NNA`), and (3) only include combinations of classes and associations appearing in previously found witnesses (`NNCA`). For example, recalling Fig. 2, after $om_1$ is found, when using the first filter `NNC`, all additional object model diff witnesses consisting of only employees, tasks, and managers, would be filtered out from the results (thus, in this case, after $om_1$ is found, no more diff witnesses will be reported).

| Name | Scope | Vars/p. vars/clauses | Alloy time (ms) | # Wit. | Plug-in time (ms) |
|---|---|---|---|---|---|
| V1 vs. V2 | 5 | 10735 / 429 / 28947 | 173 + 27 | 20 | 536 |
| – rev. – | 5 | 10735 / 429 / 28932 | 166 + 17 | 20 | 592 |
| V1 vs. V2 | 10 | 42590 / 1544 / 120542 | 1370 + 68 | 20 | 2476 |
| – rev. – | 10 | 42590 / 1544 / 120512 | 1250 + 56 | 20 | 2338 |
| V2 vs. V3 | 5 | 10947 / 429 / 29522 | 172 + 31 | 0 | 206 |
| – rev. – | 5 | 10947 / 429 / 29523 | 171 + 30 | 0 | 206 |
| V2 vs. V3 | 10 | 43442 / 1544 / 123257 | 1344 + 109 | 20 | 2761 |
| – rev. – | 10 | 43442 / 1544 / 123258 | 1422 + 97 | 20 | 2432 |
| V3 vs. V4 | 5 | 46347 / 1562 / 124413 | 1102 + 125 | 20 | 2135 |
| – rev. – | 5 | 46347 / 1562 / 124368 | 1093 + 219 | 20 | 2622 |
| V3 vs. V4 | 10 | 120807 / 3997 / 331983 | 5583 + 812 | 20 | 9821 |
| – rev. – | 10 | 120807 / 3997 / 331903 | 5617 + 384 | 20 | 11103 |
| V4 vs. V5 | 5 | 33380 / 1144 / 91631 | 678 + 173 | 20 | 1791 |
| – rev. – | 5 | 33380 / 1144 / 91617 | 674 + 66 | 20 | 1660 |
| V4 vs. V5 | 10 | 93995 / 3199 / 263066 | 4016 + 780 | 20 | 8241 |
| – rev. – | 10 | 93995 / 3199 / 263042 | 4047 + 291 | 20 | 7051 |

**Table 2.** Results from computing diff witnesses for the library example, using different scopes. Each example is reported twice, computing the differences in both directions. Columns are the same as the ones in Table 1.

Table 3 shows the results of applying our filtering mechanisms to the diff witnesses computation of *cddiff*. We report on applying the filters to the examples shown in Sect. 2 and to the library example (the same examples considered in Sect. 5.3). Note that the cases where there are no diff witnesses are omitted from the table because they are irrelevant for the filtering issue.

In all cases we first ran *cddiff* without the filters and saw that it produces at least 20 diff witnesses. Then we ran it again, each time with a different filter. The table shows the effectiveness of the filters in significantly reducing the number of witnesses. Moreover, the remaining witnesses are guaranteed to be rather different from one another and thus interesting for the engineer. Note, however, that the effectiveness of these filters depends, to a certain extent, on the order in which Alloy finds the instances, which, unfortunately, is undefined. Thus, for example, we may end up with a different set of witnesses each time we run *cddiff* with the same two CDs as input. Also, a larger scope does not guarantee that we are left with more witnesses after filtering (see, e.g., in the last section of Table 3, increasing the scope from 5 to 10 for Lib. V4 vs. V5 rev. reduced the number of witnesses that passed the filters from 3/2/4 to 3/1/4).

The filters described above and are reported on in the experiments can be considered *incremental* or *online* filters, because they are applied to the results online, as they are found during the computation. Alternatively, we may suggest *static* filters, which take the complete set of all the computed diff witnesses (up to the given scope), apply a classification based on some criteria, and then output a representative witness from each equivalence class. For example, a possible criteria for classification may be the set of classes represented in the diff witness object model. Two diff witnesses would be considered equivalent if they contain object instances from exactly the same set of classes. This would ensure variability in the set of representatives that is included in the final output. Note that the witnesses provided by the incremental filter NNC, which we described above, can all be viewed as representatives of different equivalence classes. However, the alternative static filter variant is better, as its output may be more complete and include representatives of additional equivalence classes.

| Name | Scope | # Wit. | # After filtering by NNC / NNA / NNCA |
|---|---:|---:|---:|
| Ex. 1 | 5 | 20 | 2 / 3 / 3 |
| Ex. 1 rev. | 5 | 20 | 3 / 2 / 4 |
| Ex. 1 | 10 | 20 | 2 / 2 / 3 |
| Ex. 1 rev. | 10 | 20 | 3 / 3 / 6 |
| Ex. 2 rev. | 5 | 20 | 2 / 1 / 3 |
| Ex. 2 rev. | 10 | 20 | 1 / 1 / 2 |
| Lib. V1 vs. V2 | 5 | 20 | 5 / 4 / 5 |
| Lib. V1 vs. V2 rev. | 5 | 20 | 4 / 2 / 4 |
| Lib. V1 vs. V2 | 10 | 20 | 2 / 2 / 3 |
| Lib. V1 vs. V2 rev. | 10 | 20 | 4 / 2 / 4 |
| Lib. V2 vs. V3 | 10 | 20 | 4 / 3 / 6 |
| Lib. V2 vs. V3 rev. | 10 | 20 | 5 / 3 / 6 |
| Lib. V3 vs. V4 | 5 | 20 | 3 / 2 / 4 |
| Lib. V3 vs. V4 rev. | 5 | 20 | 4 / 1 / 4 |
| Lib. V3 vs. V4 | 10 | 20 | 4 / 3 / 4 |
| Lib. V3 vs. V4 rev. | 10 | 20 | 4 / 2 / 6 |
| Lib. V4 vs. V5 | 5 | 20 | 3 / 2 / 4 |
| Lib. V4 vs. V5 rev. | 5 | 20 | 3 / 2 / 4 |
| Lib. V4 vs. V5 | 10 | 20 | 3 / 2 / 4 |
| Lib. V4 vs. V5 rev. | 10 | 20 | 3 / 1 / 4 |

**Table 3.** Results from applying filters to the examples shown in Sect. 2 and to the library example (the same examples considered in Sect. 5.3). Note that the cases where there are no diff witnesses are omitted from the table, because these are not relevant to the filters.

The use of the different filters in our plug-in is optional. Further evaluation of the effectiveness of these filters and the development of additional ones are left for future work.

### 6.2 Abstraction

Abstraction, a fundamental concept in model-driven engineering, has an important role in the context of CD comparisons. Specifically, two models may be equivalent at one level of abstraction but different in a less abstract level. Thus, the level of abstraction of interest should be defined by the engineer applying the comparison, who may be aware that the models at hand differ at a certain detailed level, but would be interested in comparing them at a higher level, where they are supposedly equivalent.

To this end, we have defined and implemented an *attribute abstraction*. With this abstraction in effect, *cddiff* ignores differences that are caused only by local changes to the attribute lists of the classes in the diagrams. That is, all class attributes of primitive or library types are abstracted away, so that two CDs whose sole difference is at the attributes level are considered equivalent. For example, in Fig. 4, if an attribute `ID` is added to the class `Employee` (in only one of the CDs) or to the abstract class `Person`, the two CDs are still considered semantically equivalent under the attribute abstraction.

The attribute abstraction becomes useful when the engineer is aware of attribute-level differences resulting from local changes, but is interested in checking for more global semantic differences, if any. Another application of this abstraction relates to performance and scope. Given two large CDs, with many classes or many attributes per class, one can start by a comparison with the abstraction in effect. If a difference is found, indeed this proves that the CDs' semantics are different. If a difference is not found, however, one has no choice but to make the comparison again with a higher scope or without the abstraction.

As a concrete example, we have compared the performance and completeness of *cddiff* with and without the attribute abstraction when running on CDs from the library example. Recall that in this example, each CD has 11 classes and the average number of attributes per class is 4. The results are shown in Table 4. On the one hand, the results show that the abstraction can reduce the size of the problem for Alloy and accelerate the computation of the diff witnesses. On the other hand, as expected, the analysis with abstraction is incomplete: in some cases it does not find all the diff witnesses that can be found without the abstraction. For example, the results for V3 vs. V4 with scope 5 show that 20 witnesses were found without abstraction, but none were found with abstraction. Interestingly, in the case of V4 vs. V5, the abstraction caused Alloy to construct an empty formula: the only differences between V4 and V5 are in some attributes and thus, without them, Alloy's formula construction and minimization was able to directly reduce the differencing predicate to false. In contrast, in the case of V2 vs. V3, the size of the formula constructed by Alloy, with or without abstraction, was the same. This happened because the differences between V2 and V3 are all *not* in the attributes, and so the optimization we use, of removing same-name attributes from same-name classes (see the end of Sect. 4.3), has the same effect as the attribute abstraction.

Defining and implementing additional abstractions to be supported by *cddiff*, e.g., an abstraction based on the composition hierarchy between classes or the containment hierarchy of packages and classes, is left for future work.

| Name | Scope | Vars/p. vars/clauses | Alloy time (ms) | # Wit. | Plug-in time (ms) |
|---|---|---|---|---|---|
| V2 vs. V3 | 5 | 10947/429/29522 | 168 + 30 | 0 | 202 |
| – w./abs. – | 5 | 10947/429/29522 | 165 + 39 | 0 | 207 |
| V2 vs. V3 | 10 | 43442/1544/123257 | 1302 + 114 | 20 | 2652 |
| – w./abs. – | 10 | 43442/1544/123257 | 1371 + 108 | 20 | 2577 |
| V3 vs. V4 | 5 | 46347/1562/124413 | 1123 + 97 | 20 | 2159 |
| – w./abs. – | 5 | 12952/486/34336 | 269 + 45 | 0 | 320 |
| V3 vs. V4 | 10 | 120807/3997/331983 | 5547 + 731 | 20 | 9508 |
| – w./abs. – | 10 | 51302/1756/143631 | 1747 + 153 | 20 | 3161 |
| V4 vs. V5 | 5 | 33380/1144/91631 | 673 + 173 | 20 | 1681 |
| – w./abs. – | 5 | 0/0/0 | 184 + 0 | 0 | 187 |
| V4 vs. V5 | 10 | 93995/3199/263066 | 4091 + 802 | 20 | 8154 |
| – w./abs. – | 10 | 0/0/0 | 1540 + 0 | 0 | 1542 |

**Table 4.** Results from computing diff witnesses for the library example, with and without the attribute abstraction. Columns are the same as the ones in Table 1.

## 7 Discussion and Future Directions

We discuss some limitations of our work and list related future work directions.

### 7.1 Bounded analysis and the small scope hypothesis

The use of Alloy, and consequently, encoding the problem of computing the diff witnesses as an instance of SAT, carries a significant price: all analysis is bounded to the user-specified scope. If a witness is found, we know the CDs' semantics are different; if no witness is found, we do not know whether the CDs have equal semantics or there still is a witness of a larger size. Recall that by size we mean the maximal number of objects in the object model. As a simple example, assuming a difference in multiplicities, between $*$ and $0..m$, a witness of size $< m$ does not exist. In this sense, the analysis is sound but incomplete. It is important to note, though, that for a given scope $k$, the analysis is sound and complete: if a witness of size $\leq k$ exists, it is found.

Nevertheless, our experience with CD, as well as an informal survey we have conducted by checking hundreds of CDs that appear in several textbooks and in different projects, e.g., the meta-model of the UML (available in [27]), showed us that while the number of classes and associations in large CDs can be high (we have seen examples of CDs with more than 100 classes), the multiplicities used on associations are typically 0..1, 1, 1..*, and *. Multiplicities that use specific numbers greater than 1 (e.g., a polygon class that has 3..* sides, a panel that has 1..10 buttons), are rather rare.

Thus, as the scope limitation is relevant mostly to the multiplicities, we adapt the *small scope hypothesis* of [13] to our problem domain, and suggest that in many cases, although the CDs involved may be large and include many classes and associations, witnesses for their differences could be rather small.

Moreover, the optimization suggested at the end of Sect. 4.3 helps us in coping with reducing the size of the problem for Alloy.

Still, given large CDs, or diagrams with no object models of small size, a symbolic technique or an abstraction/refinement approach may be recommended and required in order to allow our analysis to scale (see also subsection 6.2). Alternatively, it may be possible to identify cases where one can formally prove that a certain scope is 'good enough', that is, it may be possible to find sufficient conditions on the two CDs that will guarantee that a bounded analysis in this case is as complete as an unbounded one. We leave these directions for future research work.

### 7.2 Integration with operation-based and syntactic differencing

Our approach to semantic differencing is state-based rather than operation-based (on the distinction between the two see [23]). That is, the input for *cddiff* consists only of the two versions of the CD, and includes no information about the edit operations, if any, that have led from the first to the second version. Some works, however, concentrate on operation-based differencing, or take the two versions and aim to reconstruct a (shortest) series of edits (additions, deletions, updates) that leads from one version to the other (see the related work discussion in Sect. 8). Moreover, our approach to differencing is semantic, while most related comparison approaches are syntactic.

Thus, it may be useful to combine syntactic and operation-based differencing with state-based semantic differencing of class diagrams. For example, one may extend the applicability of semantic differencing in comparing diagrams whose elements have been renamed or moved in the course of evolution, by applying a syntactic matching (see, e.g., [9]) before computing a semantic differencing. This would result in a mapping plus a set of diff witnesses. As another example, one may find ways to use information extracted from syntactic differencing as a means to localize and thus improve the performance of semantic differencing computations.

We leave these ideas for future work.

## 8 Related Work

We discuss related work in the area of CD formal semantics and analyses and in the area of model and program comparisons.

### 8.1 CD formal semantics and analysis

Class diagrams are part of the UML standard and are widely used for the modeling of the structure of object-oriented systems, in particular in model-driven design and development setups. As such, many researchers have discussed the semantics of class diagrams and considered related analysis questions.

A number of works consider various analysis problems related to class diagrams (see, e.g., [6, 20, 33]). These include the finite satisfiability problem, the consistency between UML models, the problem of class equivalence, the identification of implicit consequences etc. Some of these works use Description Logic (DL) as their underlying formalism, some use linear programming methods, while others include no implementation but present theoretical results about the decidability and complexity of the problems at hand. In contrast, we consider the specific problem of semantic comparison and the generation of diff witnesses. We provide a solution, in a bounded scope, using a reduction to an Alloy module and its analysis with a SAT solver.

Some previous works consider the use of Alloy for the analysis of class diagrams (see, e.g., [4, 32]). These work focus on the formal definition of the transformation of a single CD to an Alloy module at the level of a meta-model and on the implementation of this transformation using a transformation language. Possible applications of the use of Alloy to analyze a given CD are not discussed in depth in these works. In contrast, as explained earlier, the input for our transformation consists of two CDs, and it produces a single Alloy module whose all instances, if any, represent the required diff witnesses. Defining and implementing our transformation using QVT or other transformation language such as ATL [15] is possible, but is outside the focus of our work.

Finally, in another paper in this conference [19] we presented modal object diagrams (MOD), as an extension of classical object diagrams, and a related verification process, which verifies a CD against an MOD specification. MOD verification is implemented using a transformation to Alloy, whose input is a CD and an MOD. It is different than the one we use here for *cddiff*.

### 8.2 Model and program comparisons

Model and program differencing, in the context of software evolution, has attracted much research efforts in recent years (see [1, 10, 17, 22, 26, 34]). In contrast to our work, almost all studies in this area, however, present syntactic differencing, at either the concrete or the abstract syntax level.

Alanen and Porres [1] describe the difference between two models as a sequence of elementary transformations, such as element creation and deletion and link insertion and removal; when applied to the first model, the sequence of transformations yields the second. Kuster et al. [17] investigate differencing and merging in the context of process models, focusing on identifying dependencies and conflicts between change operations. Engel et al.[10] present the use of a model merging language to reconcile model differences. Comparison is done by identifying new/old MOF IDs and checking related attributes and references recursively. Results include a set of additions and deletions, highlighted in a Diff/Merge browser. Mehra et al. [22] describe a visual differentiation tool where changes are presented using editing events such as add/remove shape/connector etc. Xing and Stroulia [34] present an algorithm for object-oriented design differencing whose output is a tree of structural changes, reporting differences in terms of additions, deletions, and moves of model elements, assisted by a set of similarity metrics. Ohst et al. [26] compare UML documents by traversing their abstract-syntax trees, detecting additions, deletions, and shifts of sub-trees.

As the above shows, some works go beyond the concrete textual or visual representation and have defined the comparison at the abstract-syntax level, detecting additions, removals, and shifts operations on model elements. However, to the best of our knowledge, no previous work considers model comparisons at the level of the semantic domain, as is done in our work.

Some works, e.g. [9, 34], use similarity-based matching before actual differencing. As our work focuses on semantics, it assumes a matching is given. Matching algorithms may be used to suggest a matching before the application of semantic differencing. The result of such an integration would be a mapping plus a set of differentiating traces.

We are aware of only a few studies of semantic differencing between programs. Jackson and Ladd [14] presented a tool that summarizes the semantic diff between two procedures in terms of observable input-output behaviors. Apiwattanapong et al. [5] presented a behavioral differencing algorithm for object-oriented programs based on an extended control-flow graph representation, and a tool called JDiff, which implements it in the context of Java. Finally, Person et al. [28] suggested to compute a behavioral characterization of a program change using a technique called differential symbolic execution. We focus on model comparison and not on program comparison. Also, while our work is somewhat similar to these works in terms of motivation, it is very different in terms of technology.

## 9 Conclusion

We presented *cddiff*, a semantic differencing operator for class diagrams. Unlike existing approaches to model's comparison, *cddiff* performs a semantic comparison and outputs a set of diff witnesses, each of which is an object model that is possible in the first CD and is not possible in the second. We have formally defined *cddiff*, described the technique to compute it, and demonstrated its application in comparing CDs within the Eclipse IDE. When applied to the version history of a given CD, *cddiff* provides a semantic insight into its evolution, which is not available in existing syntactic approaches.

We have implemented *cddiff* and applied it to several examples. We have extended the basic *cddiff* technique with a filtering mechanism that filters out 'uninteresting witnesses' and reports a more succinct yet informative set of witnesses to the engineer. We have also extended the basic *cddiff* technique with an attribute abstraction mechanism. This abstraction becomes useful when the engineer is aware of attribute-level differences resulting from local changes, but is interested in checking for more global semantic differences, if any. It is also useful in addressing the scope limitation and in improving *cddiff* performance.

We discussed a number of challenges and directions for future work in Sect. 7, including the development of heuristics to improve the performance of *cddiff* and allow it to scale. An interesting future work is to extend *cddiff* with support for abstraction beyond the attribute abstraction we have already defined and implemented. Another direction for future work is the integration of *cddiff* with existing approaches to matching and syntactic differencing, in particular as a means to improve its performance and the usefulness of its results to the engineers. The usefulness of *cddiff* to engineers, in particular in comparison with existing syntactic approaches, should be empirically evaluated.

Finally, in a recent paper [18] we have described our more general vision on semantic model differencing. Thus, *cddiff* is part of a larger project [31], aiming to apply the idea of semantic differencing and the computation of diff witnesses to other modeling languages, including, e.g., activity diagrams, statecharts, and feature diagrams. We hope to report on our work in these directions in future papers.

**Acknowledgments** We are grateful to Martin Schindler for defining the MontiCore language support for CDs. We thank Smadar Szekely and Guy Weiss for their expert advice on Eclipse plug-in development. We thank Eric Bodden, David Lo, and the anonymous reviewers for comments on a draft of this paper.